\documentclass[prl,twocolumn,floatix,amssymb,showpacs,amsmath,superscriptaddress]{revtex4-1}
\usepackage{graphicx}
\usepackage{epsfig,color}
\usepackage{amsmath,amssymb}
\begin{document}

\newcommand{\ket}[1]{\ensuremath{\left|{#1}\right\rangle}}
\newcommand{\bra}[1]{\ensuremath{\left\langle{#1}\right|}}
\newcommand{\quadr}[1]{\ensuremath{{\not}{#1}}}
\newcommand{\quadrd}[0]{\ensuremath{{\not}{\partial}}}
\newcommand{\slpar}{\partial\!\!\!/}
\newcommand{\gtrescero}{\gamma_{(3)}^0}
\newcommand{\gtresuno}{\gamma_{(3)}^1}
\newcommand{\gtresi}{\gamma_{(3)}^i}

\title{Quantum Simulation of Noncausal Kinematic Transformations}

\date{\today}

\author{U. Alvarez-Rodriguez}
\affiliation{Department of Physical Chemistry, University of the Basque Country UPV/EHU, Apartado 644, 48080 Bilbao, Spain}
\author{J. Casanova}
\affiliation{Department of Physical Chemistry, University of the Basque Country UPV/EHU, Apartado 644, 48080 Bilbao, Spain}
\author{L. Lamata}
\affiliation{Department of Physical Chemistry, University of the Basque Country UPV/EHU, Apartado 644, 48080 Bilbao, Spain}
\author{E. Solano}
\affiliation{Department of Physical Chemistry, University of the Basque Country UPV/EHU, Apartado 644, 48080 Bilbao, Spain}
\affiliation{IKERBASQUE, Basque Foundation for Science, Alameda Urquijo 36, 48011 Bilbao, Spain}

\begin{abstract}
We propose the implementation of Galileo group symmetry operations or, in general, linear coordinate transformations, in a quantum simulator. With an appropriate encoding, unitary gates applied to our quantum system give rise to Galilean boosts or  spatial and time parity operations in the simulated dynamics. This framework provides us with a flexible toolbox that enhances the versatility of quantum simulation theory, allowing the direct access to dynamical quantities that would otherwise require full tomography. Furthermore, this method enables the study of noncausal kinematics and phenomena beyond special relativity in a quantum controllable system.  \end{abstract}

\maketitle

Quantum simulations consist in the intentional reproduction of a quantum dynamics on another quantum system that is, typically, more controllable~\cite{Feynman82}. They promise to revolutionize computing technologies allowing us to solve otherwise intractable problems with minimal experimental resources~\cite{Lloyd96}.  Several physical models have already been proposed for quantum simulations: quantum phase transitions \cite{Greiner02}, spin models \cite{Jane, Porras, Friedenauer08,Kim10,SchmidtKaler,Wunderlich}, quantum chemistry \cite{Alan,Lanyon10}, particle statistics including anyons~\cite{Pachos,Matthews}, many-body systems with Rydberg atoms~\cite{Weimar10}, quantum relativistic systems \cite{Lamata07, Gerritsma1,Casanova1,Gerritsma2, Lamata11,CasanovaQFT,Weitz,Szameit,Reznik}, interacting fermion~\cite{Casanova12} and fermion-boson~\cite{Mezzacapo12PRL,ShiCirac} models, Majorana fermions~\cite{Lutchyn,Mezzacapo12}, the quantum Rabi model in superconducting qubits~\cite{Ballester12}, and relativistic quantum mechanics in circuit QED~\cite{Pedernales13}. It is known that the computational power of a quantum simulator may overcome that of classical computers. However, the set of operations that we can apply in the former is restricted compared with the versatility of the latter. For example, a wide set of unphysical but computable operations, while formally implementable with universal quantum computers, are not accessible to current quantum simulators.

A quantum simulation can be seen as a process in which a quantum system is forced to behave according to a given mathematical model, closely reproducing its dynamics. At the same time, the simulator has a  dynamics governed by the fundamental laws of nature. In this sense, we wonder whether quantum simulators may encode processes violating their internal operating rules. In Ref.~\cite{Casanova11}, we gave a first example showing how to implement quantum simulations of phenomena beyond quantum physics. Consequently, a natural question arises: is it possible to simulate processes violating special relativity in a quantum device respecting it?
 
In this Letter, we propose a formalism that allows the implementation of Galilean boosts and, in general,  coordinate transformations, as spatial or time parity operations, at any evolution time of a quantum simulation. This is significant to increase the versatility of quantum simulators, enabling the change of reference frame {\it in situ} during an experiment. The ability of generating these computable operations allows us to obtain correlations between different reference frames. These correlations include, among others, relevant physical quantities as propagators and self-correlation functions,  that would otherwise require full state tomography. Moreover, one could also test and analyze the ultimate limits of a quantum simulator, exploring the exciting possibility of implementing noncausal kinematics as, e.g., instantaneous translations or boosts in a controllable quantum platform. This kind of formalism may also give the capability to probe the boundary between physical and unphysical evolution. We will present the proposed method in the context of linear coordinate transformations, where the Galileo group is included. Moreover, this proposal also allows the implementation of nonlinear coordinate transformations, as is the case of accelerated frames.

Let us first consider a linear transformation between spacetime coordinates $(t, x) \rightarrow (t', x')$, i.e.,  $x'_i = \sum_{j=0}^1\alpha_{i  j} x_j$ with $(x_0, x_1) = (t, x)$, and the condition $\alpha_{01}=0$, assuring  that $t'$ is not mixed  with $x$. Here, the coefficients $\alpha_{i  j}$ determine the kind of transformation that the quantum simulator will be able to reproduce at any evolution time of the dynamics. For instance, this could be a spatial $(t,x)\rightarrow (t,\alpha x)$ or time $(t,x)\rightarrow (\alpha t, x)$ dilation, coinciding with spatial or time parity operations for $\alpha = -1$. Another example may be a Galilean boost $(t,x) \rightarrow (t, x-vt)$, where $v$ is the relative velocity between two different reference frames. We point out that the broad set of transformations with $\alpha_{01}=0$ represents the most general linear coordinate mappings implementable in a quantum simulator without prior knowledge of the evolved dynamics, as we explain below.

To illustrate the formalism, we consider now a basic dynamics governed by the equation $i\partial_t \psi = -i\partial_x\psi$, where we assume natural units $(\hbar=c=1)$. This is a $1+1$ Dirac equation for a massless particle where, for simplicity, we have traced out the internal degrees of freedom. This dynamics takes place in the Hilbert space ${\cal L}^2$ that we will call the {\it simulated space}. Our aim is to realize the above mentioned coordinate transformation during the unitary evolution, in such a way that our quantum state $\psi(x,t)$ is mapped instantaneously onto $\psi(x',t')$. Notice that this transformation is noncausal and, in general, violates special relativity. Nevertheless, we will show that we are able to implement this unphysical mapping in an enhanced quantum simulator.

The possibility of implementing this kind of  operations in a quantum simulator arises from the fact that any wave function $\psi$ and operator $\theta$ can always be expressed as
\begin{eqnarray}
\psi(x,t) &=& \frac{1}{2}\bigg\{\big[ \psi(x,t) + \psi(x', t')\big]  + \big[ \psi(x,t) - \psi(x', t')\big]\bigg\}, \nonumber\\
\theta(x,t) &=& \frac{1}{2}\bigg\{\big[ \theta(x,t) + \theta(x', t')\big]  + \big[ \theta(x,t) - \theta(x', t')\big]\bigg\}.\nonumber\\ 
\end{eqnarray}
Correspondingly, for the particular case $\theta = \partial_{t,x}$, the time and spatial derivative operators are $\partial_{t,x} = \frac{1}{2}[\partial_{t,x} + \partial_{t',x'}] + \frac{1}{2}[\partial_{t,x}-\partial_{t',x'}]$.
 With these mappings, we can write the dynamical equation, $i\partial_t\psi~=~-i\partial_x\psi$, in terms of its even $(e)$ and odd $(o)$  components as follows,
\begin{equation}\label{1+1eo}
i(\partial^{e}_t + \partial^{o}_t)(\psi^{e} + \psi^{o}) = -i(\partial^{e}_x + \partial^{o}_x)(\psi^{e} + \psi^{o}),
\end{equation}
where $\psi^{e, o} = \frac{1}{2}\big[ \psi(x,t) \pm \psi(x', t')\big]$,~$\partial_{t,x}^{e, o} =  \frac{1}{2}[\partial_{t,x}~\pm~\partial_{t',x'}]$.
Now, we define a spinor $\Psi(x, t)$ which belongs to $\mathbb{C}^2\otimes {\cal L}^2$, that we call  the {\it enlarged space}, according to $\Psi = (\psi^e, \psi^o)^T$,  where $T$ is the transpose operation. The spinor $\Psi$ is related to $\psi$ through the expression $\psi(x, t) = (1, 1)\Psi$.  An action as $\psi(x, t) \rightarrow \psi(x', t')$ is, in general, forbidden in the simulated space because it would violate special relativity. For instance, the transformation $(t, x) \rightarrow (t, -x)$ is an operation that produces an instantaneous translation and inversion of a  wavepacket centered in an  average value $x_0$ to an average value $-x_0$. In this sense, it would violate causality and the no-signalling condition for large enough $x_0$. For example, this action applied on the wave function $\psi(x,t)=\exp[-(x-x_0)^2/\gamma_x^2]$,  where $\gamma_x$ corresponds to the Gaussian width, produces $\exp[-(x+x_0)^2/\gamma_x^2]$. This instantaneous operation is forbidden in the simulated space by  special relativity. Nevertheless, the spinor $\Psi'$ in the enlarged space, corresponding to $\psi(x', t')$, is just $\sigma_z \Psi$, i.e., $\psi(x', t') = (1, 1)\sigma_z\Psi(x, t)$. This means that a physical action like $\sigma_z$, acting on the enlarged space, gives rise to a physically-forbidden action on the wave function in the simulated space.

The dynamical equation for $\Psi(x, t)$ can be obtained from Eq.~(\ref{1+1eo}) separating its even and odd components,  giving rise to
\begin{equation}\label{matrixeq}
i\left(\begin{array}{cc}
\partial_t^e& \partial_t^o\\
\partial_t^o& \partial_t^e\\
\end{array}\right)\left(\begin{array}{c}
\psi^e\\
\psi^o
\end{array}\right) = -i \left(\begin{array}{cc}
\partial_x^e& \partial_x^o\\
\partial_x^o& \partial_x^e\\
\end{array}\right)\left(\begin{array}{c}
\psi^e\\
\psi^o
\end{array}\right).
\end{equation}
We can write $\partial_{t, x}^{e, o}$ in terms of $\partial_t$ and $\partial_x$ as follows,
\begin{eqnarray}
\partial^{e, o}_t &=& \frac{1}{2} \big[\partial_t \pm \partial_{t'}  \big]  = \frac{1}{2} \big[\partial_t \pm \partial_{\alpha_{0 0} t}  \big]\nonumber\\ &=& \frac{1}{2}\bigg[\partial_t + \frac{1}{\alpha_{1 1}\alpha_{0 0} } \big(\pm \alpha_{1 1}\partial_t \mp \alpha_{1 0}\partial_x\big)\bigg] ,
\end{eqnarray}

and also
 \begin{eqnarray}
\partial^{e, o}_x &=& \frac{1}{2} \big[\partial_x \pm \partial_{x'}  \big]  = \frac{1}{2} \big[\partial_x \pm \partial_{(\alpha_{1 1} x + \alpha_{1 0} t)}  \big]\nonumber \\ &=& \frac{1}{2}\bigg[\partial_x+ \frac{1}{\alpha_{1 1}\alpha_{0  0}}(\pm \alpha_{0 0} \partial_x) \bigg] .
\end{eqnarray}
We can substitute these expressions in Eq.~(\ref{matrixeq}) in order to obtain a Schr\"odinger equation for $\Psi$,
\begin{equation}\label{Psit}
i\partial_t \Psi = -i \big[\tilde{\alpha}_1 I + \tilde{\alpha}_2 \sigma_x \big]\partial_x\Psi,
\end{equation}
with $\tilde{\alpha}_{1,2} = \frac{\alpha_{1 1} \pm \alpha_{0 0} \mp \alpha_{1 0}}{2(\alpha_{1 1})}$. In order to compute the dynamics associated with Eq.~(\ref{Psit}), one just has to define the initial condition for $\Psi$, i.e., $\Psi(x,0) = \frac{1}{2} [ \psi(x,0) + \psi \big(x'(x,0), t'(x,0) \big) , \psi(x,0) - \psi \big( x'(x,0), t'(x,0) \big) ]^T$. Notice that this state is completely determined by the initial condition $\psi(x, 0)$ of the dynamics in the simulated space only if $t'(x,0)=0$. Otherwise, prior knowledge of the full spatial and time dependence of the wavefunction $\psi(x,t)$ is needed to determine $\Psi(x,0)$. This is a circular argument given that the time dependence of $\psi$ is what we want to compute in the quantum simulation. We point out that the kind of initial condition with $t'(x,0)=0$ does not include Lorentz boosts in the genuinely relativistic regime. This is due to the fact that, in the case of relativistic Lorentz boosts, $t'(x,t)$ is a linear superposition of $x$ and $t$. This means that, in general, the condition with vanishing $t'$ is not fulfilled. Accordingly, the set  of allowed linear transformations without prior knowledge of the full wavefunction is given by  $t'=\alpha_{00} t, x'=\alpha_{10} t + \alpha_{11} x$, including among others the Galileo group. Moreover, extensions to nonlinear coordinate transformations, as $t'=\alpha_{00} t^k, x'=f(x,t)$, with $f(x,t)$ a certain function of $x$ and $t$, can also be considered. This is  because they have valid initial conditions as well,  $\psi(x'(x,0),t'(0))=\psi(f(x,0),0)$, i.e., conditions that are  univocally determined by the knowledge of $\psi(x, 0)$. 

Equation~(\ref{Psit}) includes  the dynamics of $\psi(x, t)$ and $\psi(x', t')$, allowing us to obtain any expectation value of both dynamics relating it to the measurement of observables in the enlarged space. This can be expressed as
\begin{eqnarray}
\langle O \rangle_{\psi(x, t)} = \langle\psi| O |\psi\rangle &=& \langle\Psi|\left(\begin{array}{c}
1\\
1\end{array}\right) O  \left(\begin{array}{cc} 1  &,1\end{array}\right)|\Psi\rangle\nonumber\\ &=& \langle\Psi| (I + \sigma_x)\otimes O |\Psi\rangle,\\
\langle O \rangle_{\psi(x', t')} = \langle\psi'| O |\psi'\rangle &=& \langle\Psi|\sigma_z\left(\begin{array}{c}
1\\
1\end{array}\right) O  \left(\begin{array}{cc} 1  &,1\end{array}\right)\sigma_z|\Psi\rangle\nonumber\\ &=& \langle\Psi| (I - \sigma_x)\otimes O |\Psi\rangle,
\end{eqnarray}
where we use $\langle x|\psi\rangle = \psi(x, t)$, $\langle x'|\psi'\rangle = \psi(x', t')$, and $\langle x|\Psi\rangle=\Psi(x, t)$.

With the same tools we can obtain information about correlations between $\psi(x, t)$ and $\psi(x', t')$ in terms of measurements of observables in the enlarged space,
\begin{eqnarray}
\langle O \rangle_{\psi(x, t), \psi(x', t')}  = \langle\psi| O |\psi'\rangle &=& \langle\Psi|\left(\begin{array}{c}
1\\
1\end{array}\right) O  \left(\begin{array}{cc} 1  &,1\end{array}\right)\sigma_z|\Psi\rangle\nonumber\\ &=& \langle\Psi| (\sigma_z - i \sigma_y)\otimes O |\Psi\rangle.
\end{eqnarray}

We point out that extending our formalism to many-body systems may be done with a systematic approach. For example, for simulating fermions with a trapped-ion simulator, it suffices to consider one ion per simulated particle by using recently developed techniques~\cite{Casanova12}. The application of our protocol to systems  of many particles~\cite{britton12} will give rise to measurements of correlations otherwise difficult to perform. This is because the process for obtaining the correlations involves the realization of two  different quantum simulations, one for $\psi$ and another one for $\psi'$. Subsequently, we have to implement a full quantum tomography of both cases and finally store all the data in a classical computer and calculate the correlations.  In general, if we are dealing with dynamics involving many particles, realizing the full tomography is demanding, while extracting the useful information can be accomplished applying our proposed ideas. 

Our method also allows us  to include nonlocal and noncausal  operations directly into the dynamics.  For example, the equation
\begin{equation}\label{pot}
i\partial_t\psi = \big[ i\sigma_x p_x \Pi_x + m\sigma_z + \sigma_x V(x) \big]\psi,
\end{equation}
where $\Pi_x\psi(x, t) = \psi(-x, t)$, contains the spatial parity operator $\Pi_x$ in the kinetic term. This means that the operation $(x, t)\rightarrow(-x,t)$, $(\alpha_{0 0}, \alpha_{0 1}, \alpha_{1 0}, \alpha_{1 1}) = (1, 0, 0, -1)$ is included in the dynamics of the simulator.

Equation (\ref{pot}) cannot be implemented directly in an experimental setup because the Hamiltonian contains an unphysical (spacetime nonlocal) operation. Nevertheless, through  the mapping $\psi(x, t) \rightarrow \Psi = (\psi(x, t) + \psi(-x, t),  \psi(x, t) - \psi(-x, t))^T  $,  encoding properly the action of $\Pi_x$ on Pauli operators, we obtain its image in an enlarged space, 
\begin{eqnarray}
i\partial_t\Psi \!=\! \bigg[p\sigma_y\otimes\sigma_x + m I\otimes\sigma_z
+ V^{e}I\otimes\sigma_x + V^{o}\sigma_x\otimes\sigma_x\bigg]\!\Psi.\nonumber
\end{eqnarray}
Here, $V(x) = V^e(x) + V^o(x)$ is decomposed in its corresponding even ($e)$ and odd ($o$) parts. We consider now the nonrelativistic regime $m \gg  \big(|\langle  p \rangle|, \ |V(x)|\big)$, and the case in which  the potential is explicitly odd, $V(x) = V^o(x)$. Then,  Eq.~(\ref{pot}) can be written in an interaction picture with respect to the mass term, and for a parity eigenstate associated to the $\pm$ eigenvalue, we have

\begin{equation}\label{sicausal}
i\partial_t\psi = \frac{\sigma_z}{2m}\bigg[ p_x^2 + (V^o)^2 \pm (\partial_xV^o) \bigg]\psi.
\end{equation}
This equation describes the causal behavior of a nonrelativistic particle under the influence of an even potential. This shows an intriguing relation between the dynamics of  Eq.~(\ref{pot}), restricted to the case of odd potentials, and the dynamics of its nonrelativistic limit given by Eq.~(\ref{sicausal}). The first one shows a noncausal behavior evolving under an odd potential, while the second one is a causal equation where the effective potential that emerges is even. 

\begin{figure}[t]\centering
\includegraphics[width=65mm]{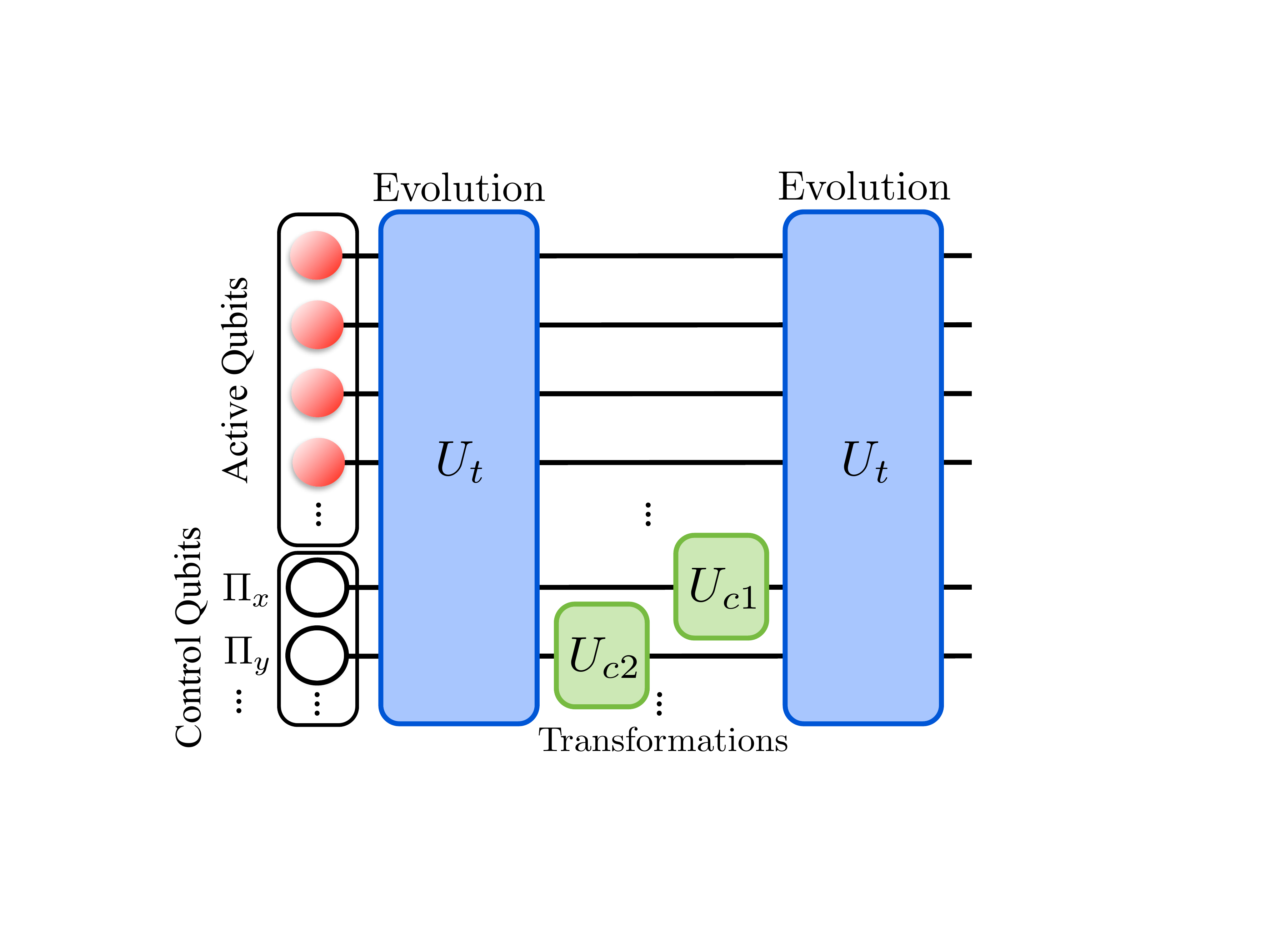}  
\caption{ Scheme of gates acting on the {\it active} (coloured circles) and {\it control} (white circles) qubits, implementable in different quantum platforms. In this example, we depict a quantum simulation process in which two or more (spacetime nonlocal) kinematic transformations $\Pi_x$, $\Pi_y$, ... have been performed at an intermediate time of the simulated dynamics. The first and fourth boxes (starting from the right), are used to generate the dynamics given by an evolution operator $U_t$. The small boxes represent local gates $U_{c1, c2}$ acting on the control qubits.}
   \label{Fig2}
\end{figure}

Another example of the transformations we can include is the time parity, $(t, x)\rightarrow(-t, x)$, $(\alpha_{0 0}, \alpha_{0 1}, \alpha_{1 0}, \alpha_{1 1}) = (-1, 0, 0, 1)$. Time is a global parameter common to all quantum particles, which is introduced in quantum mechanics through the Schr\"odinger equation $i\partial_t\psi = H \psi$.  We include this operation in our formalism through the equality
\begin{equation}\label{time1}
\psi(x,t) = \frac{1}{2}\bigg\{\big[\psi(x, t) + \psi(x, -t)\big] + \big[\psi(x, t) - \psi(x, -t)\big]\bigg\}.
\end{equation}
Notice that $\psi(x, -t)$ can be expressed as
$\Pi_t \psi(x, t) = \frac{1}{2}[\psi(x, t) + \psi(x, -t)] - \frac{1}{2}[\psi(x, t) - \psi(x, -t)]$, where $\Pi_t$ is the parity operator. Accordingly, performing this mapping amounts to introducing a minus sign in the second term of Eq.~(\ref{time1}). 
The operation $\Pi_t$ cannot be realized in the real world because it implies instantaneous travel from the future to the past for a set of particles. Nevertheless, it  can be implemented using an enlarged space as follows. The corresponding Schr\"odinger equation for $H\neq H(t) $ can be written as
\begin{equation}
i\partial_t (\psi^e + \psi^o ) = H^e(\psi^e + \psi^o ),
\end{equation}
where $\psi^e = \frac{1}{2}[\psi(x, t) + \psi(x, -t)]$, $\psi^o = \frac{1}{2}[\psi(x, t) - \psi(x, -t)]$ and $H^e = H$, given that the Hamiltonian does not depend on time. This equation can be separated into even and odd parts and expressed in an enlarged space,
\begin{equation}\label{parityeq}
i \partial_t\left(\begin{array}{c}
\psi^e\\
\psi^o
\end{array}\right) = \sigma_x\otimes H^e \left(\begin{array}{c}
\psi^e\\
\psi^o
\end{array}\right).
\end{equation}
Notice that $\partial_t \psi^e$ is odd with respect to the transformation $t \rightarrow -t$, and $\partial_t \psi^o$ is even. Here, the relation between spinors in the simulated and enlarged spaces is again $\psi = (1, 1) \Psi$ with $\Psi = (\psi^e, \psi^o)^T$. 
Now, at any time of the evolution, we can apply a time parity in the enlarged space multiplying the state at time $t$ by the gate $\sigma_z$. This is due to the fact that $\psi(x,-t) = (1,1) \sigma_z\Psi(x, t)$. In this case, to prepare the initial state in the enlarged space is especially easy given that
\begin{equation}
\Psi(x, t=0) = \left(\begin{array}{c}
\psi(x,t=0)\\
0
\end{array}\right) = \left(\begin{array}{c}
1\\
0
\end{array}
\right)\otimes\psi(x, t=0).
\end{equation}
This is because at $t=0$, $\psi^e= \frac{1}{2}[\psi(x, 0) + \psi(x, 0)]=\psi(x,0)$ and $\psi^o= \frac{1}{2}[\psi(x, 0) - \psi(x, 0)]=0$. 

The inclusion of the time parity operation allows us to encode the expectation value of the propagator $e^{-iHt}$  in two observables in the enlarged space. We point out that computing this average value requires in general full tomography with alternative approaches. To accomplish this task, we first   evolve the initial wavefunction $\Psi(x, t=0)$ under the Hamiltonian $\sigma_x\otimes H_e$ of  Eq.~(\ref{parityeq}), generating the state $\Psi(x, t)=\frac{1}{2}(\psi(x,t) + \psi(x, -t), \psi(x,t) - \psi(x, -t))^T$. Later, we apply $e^{-iI\otimes H^{e} \Delta}$ to  $\Psi(x, t)$, producing
\begin{equation}
\tilde{\Psi}=e^{-iI\otimes H^{e} \Delta}\Psi(x, t)\bigg|_{\Delta=t} =\frac{1}{2}\left(\begin{array}{c}
\psi(x, 2t) + \psi(x, 0)\\
\psi(x, 2t)  - \psi(x, 0)
\end{array}\right).\\
\end{equation}

Thus, according to the following equivalence,
\begin{eqnarray}\label{propagator}
\langle e^{-i2tH} \rangle&=&\langle\psi(x, 0)|\psi(x,2t)\rangle=\langle\tilde\Psi|\sigma_z\left(\begin{array}{c}
1\\
1
\end{array}\right) \left(\begin{array}{cc} 1  &,1\end{array}\right)|\tilde\Psi\rangle\nonumber\\
&=&\langle\tilde\Psi|(\sigma_z+i\sigma_y)|\tilde\Psi\rangle,
\end{eqnarray}
the measurement of $\sigma_z$ and $\sigma_y$ in the enlarged space  will provide us  with the  expectation value of the propagator. In some cases, it is possible to relate the expectation value of Eq.~(\ref{propagator}) with self-correlation functions. For example, for spin systems one can write 
\begin{eqnarray}\label{selfcorrelation}
\langle e^{-i2tH} \rangle&=&\langle\psi(s_i, t=0 )| e^{-i2tH}|\psi(s_i, t=0 )\rangle \\
&=&\langle\psi'(s_i, t=0 )|\sigma_j  e^{-i2tH}  \sigma_j |\psi'(s_i, t=0 )\rangle,\nonumber
\end{eqnarray}
where $s_i$ refers to the spin degrees of freedom,  $|\psi'(s_i, t=0 )\rangle = \sigma_j|\psi(s_i, t=0 )\rangle$, and $\sigma_j$ corresponds to $\sigma_x, \sigma_y, \sigma_z$ for $j = 1, 2, 3$, or, in general, to any Hermitian linear combination of them. In cases in which $\{H, \sigma_j \}=0$, it is possible to write the last line of Eq.~(\ref{selfcorrelation}) as
\begin{equation}
\langle\psi'(s_i, t=0 )| \sigma_j(-t/2)  \sigma_j(t/2) |\psi'(s_i, t=0 )\rangle,
\end{equation}
with $\sigma_j(\pm t/2) = \exp{\left(\mp\frac{it}{2}H\right)} \sigma_j \exp{\left(\pm\frac{it}{2}H\right)}$. This corresponds to a self-correlation function that would require in general full tomography to be computed.

Our protocol can be generalized in order to include the possibility of performing several  transformations in a quantum simulation on one or many particles. For instance, given a wave function $\psi(x, y)$, where $x$ and $y$  are independent coordinates that may represent the position of two particles in one dimension or the coordinates of one particle in two dimensions, we have $\psi(x, y)= \psi_{e, e} + \psi_{e, o} + \psi_{o, e} + \psi_{o, o}$, where 
\begin{eqnarray}\label{decomp}
\psi_{i, j}  & = & \frac{1}{4}\bigg\{\big[ \psi(x, y)  +(-1)^i  \psi(-x, y)\big] \nonumber\\&& +(-1)^j   \big[ \psi(x, -y)  + (-1)^i   \psi(-x, -y)\big]  \bigg\},
\end{eqnarray} 
and $i,j=$$\{e\equiv 0$,$o\equiv 1\}$.
We then consider the spinor $\Psi=(\psi_{e, e}, \psi_{e, o}, \psi_{o, e}, \psi_{o, o} )^T$, that is related to $\psi(x,y)$ through $\psi(x, y) = (1,1,1,1) \Psi$. The decomposition of Eq.~(\ref{decomp}) allows to generate easily operations of spatial parity in the $x$ and $y$ coordinates, $\Pi_x \psi(x, y) = \psi(-x, y)$, $\Pi_y\psi(x, y) = \psi(x, -y)$, $\Pi_x \Pi_y \psi(x, y) = \psi(-x, -y)$. This is achieved applying just local gates to the state $\Psi$  in the enlarged space, $\Pi_x \psi(x, y) = (1, 1, 1, 1) \left(\sigma_z\otimes I\right) \Psi$,
$\Pi_y \psi(x, y) = (1, 1, 1, 1) \left(I \otimes \sigma_z\right) \Psi$, $\Pi_x\Pi_y \psi(x, y) = (1, 1, 1, 1) \left(\sigma_z \otimes \sigma_z\right) \Psi$.
In general, each inclusion of a new symmetry transformation in the quantum simulation amounts to doubling the Hilbert space. In different quantum optical implementations, like trapped-ion setups~\cite{LeibfriedEtAl} and superconducting qubits~\cite{Wilhelm}, this implies adding another qubit encoding the proposed symmetry, see Fig.~\ref{Fig2}. The interactions that appear due to the inclusion of these symmetries, involving tensor products of Pauli matrices, are efficiently implementable in a digital quantum simulator with recently developed techniques~\cite{Casanova12, Lanyon11}. 

In summary, we have explored the limits of quantum simulations via a formalism performing linear coordinate transformations during a simulated quantum dynamics. Among other features, we may compute spin temporal correlation functions without performing full tomography. Moreover, our method allows us to measure directly correlations between different reference frames. Finally, we may also study noncausal kinematics in a system respecting the laws of quantum physics and relativity. We point out that these fundamental concepts and formalism may be implemented in a wide variety of platforms, as, e.g., trapped ions, superconducting qubits, cold atoms, and integrated quantum photonics.

We thank  I. L. Egusquiza for interesting discussions. The authors acknowledge funding from Basque Government  BFI-2012-322 and IT472-10 grants, EC Marie Curie IEF grant, Spanish MINECO FIS2012-36673-C03-02, UPV/EHU UFI 11/55, SOLID, CCQED, PROMISCE, and SCALEQIT European projects.


\begin{thebibliography}{10}

\bibitem{Feynman82}
R. P. Feynman,  Int. J. Theor. Phys. \textbf{21}, 467 (1982).

\bibitem{Lloyd96}
S. Lloyd, Science {\bf 273}, 1073 (1996).

\bibitem{Greiner02} M. Greiner, O. Mandel, T. Esslinger, T. W. H\"ansch and I. Bloch,  Nature {\bf 415}, 39 (2002).

\bibitem{Jane}  E. Jan\'e, G. Vidal, W. D\"ur, P. Zoller, and J. I. Cirac, Quant. Inf. and Comp. {\bf 3}, 15 (2003).

\bibitem{Porras} D. Porras and J. I. Cirac, Phys. Rev. Lett. {\bf 92}, 207901 (2004).

\bibitem{Friedenauer08} A. Friedenauer, H. Schmitz, J. T. Glueckert, D. Porras,  and T. Sch\"atz, Nature Phys. {\bf 4}, 757 (2008).

\bibitem{Kim10} K. Kim, M.-S. Chang, S. Korenblit, R. Islam, E. E. Edwards, J. K. Freericks, G.-D. Lin, L.-M. Duan and C. Monroe, Nature {\bf 465}, 590 (2010).

\bibitem{SchmidtKaler} J. Welzel, A. Bautista-Salvador, C. Abarbanel, V. Wineman-Fisher, C. Wunderlich, R. Folman, and F. Schmidt-Kaler, Eur. Phys. J. D {\bf 65}, 285 (2011).

\bibitem{Wunderlich} M. Johanning, A. F. Var\'on, and C. Wunderlich, J. Phys. B {\bf 42}, 154009 (2009).

\bibitem{Alan} A. Aspuru-Guzik, A. D. Dutoi, P. J. Love, M. Head-Gordon, Science {\bf 309}, 1704 (2005).

\bibitem{Lanyon10} B. P. Lanyon, J. D. Whitfield, G. G. Gillet, M. E. Goggin, M. P. Almeida, I. Kassal, J. D. Biamonte, M. Mohseni, B. J. Powell, M. Barbieri, A. Aspuru-Guzik and A. White,  Nature Chem. {\bf 2}, 106 (2010).

\bibitem{Pachos} J. K. Pachos, W. Wieczorek, C. Schmid, N. Kiesel, R. Pohlner, H. Weinfurter, New J. Phys. {\bf 11}, 083010 (2009).

\bibitem{Matthews} J. C. F. Matthews, K. Poulios, J. D. A. Meinecke, A. Politi, A. Peruzzo, N. Ismail, K. W\"orhoff, M. G. Thompson, J. L. O'Brien, Sci. Rep. {\bf 3}, 1539 (2013).

\bibitem{Weimar10} H. Weimer, M. M\"uller, I. Lesanovsky, P. Zoller and H. P. B\"uchler, Nature Phys. {\bf 6}, 382 (2010). 

\bibitem{Lamata07} L. Lamata, J. Le\'on, T. Sch\"{a}tz, and E. Solano, Phys. Rev. Lett. {\bf 98}, 253005 (2007).

\bibitem{Gerritsma1} R. Gerritsma, G. Kirchmair, F. Z\"{a}hringer, E. Solano, R. Blatt, C. F. Roos,  Nature {\bf 463}, 68 (2010).

\bibitem{Casanova1} J. Casanova, J. J. Garc\'{\i}a-Ripoll, R. Gerritsma, C. F. Roos, E. Solano, Phys. Rev. A {\bf 82}, 020101 (2010).

\bibitem{Gerritsma2} R. Gerritsma, B. P. Lanyon, G. Kirchmair, F. Z\"{a}hringer, C. Hempel, J. Casanova, J. J. Garc\'{\i}a-Ripoll, E. Solano, R. Blatt, and C. F. Roos, Phys. Rev. Lett. {\bf 106}, 060503 (2011).

\bibitem{Lamata11}
L. Lamata, J. Casanova, R. Gerritsma, C. F. Roos, J. J. Garc\'{\i}a-Ripoll, and E. Solano, New J. Phys. {\bf 13}, 095003 (2011).

\bibitem{CasanovaQFT}
J. Casanova, L. Lamata, I. L. Egusquiza, R. Gerritsma, C. F. Roos, J. J. Garc\'{\i}a-Ripoll, and E. Solano, Phys. Rev. Lett. {\bf 107}, 260501 (2011). 

\bibitem{Weitz} 	T. Salger, C. Grossert, S. Kling, M. Weitz, Phys. Rev. Lett. {\bf 107}, 240401 (2011).

\bibitem{Szameit} F. Dreisow, M. Heinrich, R. Keil, A. T\"unnermann, S. Nolte, S. Longhi, A. Szameit, Phys. Rev. Lett. {\bf 105}, 143902 (2010).

\bibitem{Reznik} E. Zohar, J. I. Cirac, B. Reznik, Phys. Rev. Lett. {\bf 109}, 125302 (2012). 

\bibitem{Casanova12} J. Casanova, A. Mezzacapo, L. Lamata, and E. Solano, Phys. Rev. Lett. {\bf 108}, 190502 (2012).

\bibitem{Mezzacapo12PRL} A. Mezzacapo, J. Casanova, L. Lamata, and E. Solano, Phys. Rev. Lett. {\bf 109}, 200501 (2012).

\bibitem{ShiCirac} V. M. Stojanovi\'c, T. Shi, C. Bruder, J. I. Cirac, 
Phys. Rev. Lett. {\bf 109}, 250501 (2012).

\bibitem{Lutchyn} R. M. Lutchyn, J. D. Sau, and S. Das Sarma, Phys. Rev. Lett. {\bf 105}, 077001 (2010). 

\bibitem{Mezzacapo12} A. Mezzacapo, J. Casanova, L. Lamata, and E. Solano, New J. Phys. {\bf 15}, 033005 (2013).

\bibitem{Ballester12} D. Ballester, G. Romero, J. J. Garc\'{\i}a-Ripoll, F. Deppe, and E. Solano, Phys. Rev. X {\bf 2}, 021007 (2012).

\bibitem{Pedernales13} J. S. Pedernales, R. Di Candia, D. Ballester, and E. Solano. Accepted in New Journal of Physics,  e-print: arXiv:1211.3953.

\bibitem{Casanova11} J. Casanova, C. Sab\'{\i}n, J. Le\'on, I. L. Egusquiza, R. Gerritsma, C. F. Roos, J. J. Garc\'{\i}a-Ripoll, and E. Solano, Phys. Rev. X {\bf 1}, 021018 (2011).

\bibitem{britton12} J. W. Britton, B. C. Sawyer, A. C. Keith, C.-C. J. Wang, J. K. Freericks, H. Uys, M. J. Biercuk,  and J. J. Bollinger, Nature {\bf  484}, 489 (2012).

\bibitem{LeibfriedEtAl} D. Leibfried, R. Blatt, C. Monroe, and D. Wineland, Rev. Mod. Phys. {\bf 75}, 281 (2003).

\bibitem{Wilhelm} J. Clarke and F. K. Wilhelm, Nature {\bf 453}, 1031 (2008).

\bibitem{Lanyon11} B. P. Lanyon, C. Hempel, D. Nigg, M. M\"uller, R. Gerritsma, F. Z\"ahringer, P. Schindler, J. P. Barreiro, M. Rambach, G. Kirchmair, M. Hennrich , P.Zoller, R. Blatt, and C. F. Roos, Science {\bf 334}, 57 (2011).


\end{thebibliography}
\end{document}